\documentclass[12pt]{article}
\usepackage{graphicx}
\usepackage{epstopdf}
\usepackage{amsmath}
\usepackage{amsfonts}
\usepackage{amssymb}
\usepackage{color}
\usepackage{mathrsfs}
\usepackage{multirow}

\usepackage{indentfirst}
\usepackage{amsthm,amscd}
\usepackage[table]{xcolor}
\usepackage{slashed}
\usepackage{dsfont}
\usepackage{bbold}

\newcommand{\be}{\begin{equation}}
\newcommand{\ee}{\end{equation}}
\newcommand{\barr}{\begin{array}}
\newcommand{\earr}{\end{array}}

\usepackage{latexsym}
\usepackage{amsthm}
\usepackage{amsmath}
\usepackage{graphicx}
\usepackage{amssymb}
\usepackage{bbm}
\usepackage{subfig}
\newcommand{\gsim}{\lower.7ex\hbox{$\;\stackrel{\textstyle>}{\sim}\;$}}
\newcommand{\lsim}{\lower.7ex\hbox{$\;\stackrel{\textstyle<}{\sim}\;$}}

\newcommand{\bea}{\begin{eqnarray}}
\newcommand{\eea}{\end{eqnarray}}

\newcommand{\comment}[1]{}

\def\e{\mathrm{e}}

\def\({\left(}
\def\){\right)}
\def\[{\left[}
\def\]{\right]}
\def\e{\begin{equation}}
\def\q{\end{equation}}
\def\m{\begin{eqnarray}}
\def\n{\end{eqnarray}}

\begin{document}

\setcounter{page}{1} \baselineskip=15.5pt \thispagestyle{empty}

\begin{flushright}
\end{flushright}
\vfil

\begin{center}

{\LARGE \bf Forecasting sensitivity on tilt of power spectrum of primordial gravitational waves after Planck satellite}
\\[0.7cm]
{Qing-Guo Huang $^\Diamond$ \footnote{huangqg@itp.ac.cn}, Sai Wang $^\Diamond$ \footnote{wangsai@itp.ac.cn}, Wen Zhao $^{\heartsuit}$ \footnote{wzhao7@ustc.edu.cn}}
\\[0.7cm]

{\normalsize { \sl $^\Diamond$ State Key Laboratory of Theoretical Physics, Institute of Theoretical Physics, Chinese Academy of Science, Beijing 100190, China}}\\
\vspace{.2cm}

{\normalsize { \sl $^\heartsuit$ CAS Key Laboratory for Researches in Galaxies and Cosmology, Department of Astronomy, University of Science and Technology of China, Chinese Academy of Sciences, Hefei, Anhui 230026, China}}\\
\vspace{.3cm}

\end{center}

\vspace{.8cm}

\hrule \vspace{0.3cm}
{\small  \noindent \textbf{Abstract} \\[0.3cm]
\noindent By taking into account the contamination of foreground radiations, we employ the Fisher matrix to forecast the future sensitivity on the tilt of power spectrum of primordial tensor perturbations for several ground-based (AdvACT, CLASS, Keck/BICEP3, Simons Array, SPT-3G), balloon-borne (EBEX, Spider) and satellite (CMBPol, COrE, LiteBIRD) experiments of B-mode polarizations. For the fiducial model $n_t=0$, our results show that the satellite experiments give good sensitivity on the tensor tilt $n_t$ to the level $\sigma_{n_t}\lesssim0.1$ for $r\gtrsim2\times10^{-3}$, while the ground-based and balloon-borne experiments give worse sensitivity. By considering the BICEP2/Keck Array and Planck (BKP) constraint on the tensor-to-scalar ratio $r$, we see that it is impossible for these experiments to test the consistency relation $n_t=-r/8$ in the canonical single-field slow-roll inflation models.
}
\vspace{0.5cm}
\hrule
\vspace{3.2cm}

\begin{flushleft}
\end{flushleft}

\vspace{18cm}

\newpage

\section{Introduction}\label{introduction}

The inflation paradigm \cite{Starobinsky:1979,Starobinsky:1980te,Guth:1980zm,Linde:1981mu,Albrecht:1982wi} predicts the primordial gravitational waves, i.e. tensor perturbations. The primordial gravitational waves can contribute to the total intensity and polarizations of cosmic microwave background (CMB) anisotropy \cite{Grishchuk:1974ny,Starobinsky:1979ty,Rubakov:1982,Crittenden:1993ni,Kamionkowski:1996zd,Kamionkowski:1996ks,Hu:1997mn}.
From the recent Planck TT,TE,EE+lowP data release \cite{Ade:2015lrj}, the upper limit on tensor-to-scalar ratio is given by $r_{0.002}<0.11$ at the $95\%$ confidence level (C.L.) by fitting the $\Lambda$CDM+r model. The B-mode polarizations contributed by primordial gravitational waves may be detectable at the range $\ell\lesssim150$. Actually, BICEP2 \cite{Ade:2014xna} has pushed the sensitivity of B-mode polarizations to be comparable with that of temperature in searching for the primordial gravitational waves. However, the polarized dust emissions make us difficult to distinguish whether the detected B-mode power comes from the primordial gravitational waves \cite{Mortonson:2014bja,Flauger:2014qra,Colley:2014nna,Cheng:2014pxa}. Recently, Planck \cite{Adam:2014bub} released the full-sky data of polarized dust emissions. Based on this, a joint analysis of the B-mode data from BICEP2/Keck Array and Planck (BKP) \cite{Ade:2015tva} yielded an upper bound $r_{0.05}<0.12$ at the $95\%$ C.L., which is compatible with the upper limit from Planck data without the B-mode polarizations.

The tilt of power spectrum of primordial tensor perturbations is used to measure the feature of the primordial gravitational waves. In the power spectrum of primordial tensor perturbations, the tensor tilt $n_t$ is defined by
\begin{equation}
\label{tensor power spectrum}
P_t(k)=rA_s\left(\frac{k}{k_p}\right)^{n_t}\ ,
\end{equation}
where $P_t(k)$ is the amplitude of power spectrum of primordial tensor perturbations at the scale $k$, $r$ denotes the tensor-to-scalar ratio at a given pivot scale $k_p$ and $A_s$ the amplitude of primordial scalar perturbations which is set as a constant in this paper. In the inflation model, the tensor tilt is generally predicted as $n_t=-2\epsilon$ \cite{Liddle:1992wi,Garriga:1999vw}. The inflation requires $\ddot{a}/a=H^2(1-\epsilon)$ where $\epsilon=\dot{H}/H^2$, and thus $-2<n_t<0$. In the canonical singe-field slow-roll inflation models, the tensor tilt is determined by the tensor-to-scalar ratio via the consistency relation, i.e. $n_t=-r/8$ \cite{Liddle:1992wi}. By considering the upper bounds on $r$, we expect the power spectrum of primordial gravitational waves to be nearly scale--invariant, i.e. $n_t\simeq0$. The current constraint on $n_t$ was given by $n_{t,0.01}=-0.76^{+1.37}_{-0.52}$ \cite{Huang:2015gka} at the $68\%$ C.L. by combining only the BKP B-mode data and the upper limit on the intensity of stochastic gravitational wave background from Laser Interferometer Gravitational-Waves Observatory (LIGO) \cite{Aasi:2014zwg}. The scale--invariant spectrum is well compatible with this constraint.

Even though at present there are no evidence for the primordial gravitational waves, several future polarization experiments might reach the sensitivity to detect the primordial gravitational waves in the coming years. As a recent analysis, Ref.~\cite{Creminelli:2015oda} forecasted that the primordial gravitational waves with theoretically motivated $r\sim2\times10^{-3}$ can be achievable by certain future experiments if the noise is reduced to $\sim1\mu K$-arcmin and the lensing B-modes reduced to $10\%$. Their forecasts are not changed significantly with respect to previous estimates \cite{Lee:2014cya}. In this paper, we study the sensitivity on the tensor tilt  $n_t$ for several future ground-based (AdvACT, CLASS, Keck/BICEP3, Simons Array, SPT-3G), balloon-borne (EBEX, Spider) and satellite (CMBPol, COrE, LiteBIRD) experiments. We are just interested in studying the nearly scale-invariant case, i.e. $n_t\simeq0$ which corresponds to a class of the simplest inflation models. Similar to Ref.~\cite{Creminelli:2015oda}, the approach of Fisher matrix is also used in our analysis. The paper is arranged as follows. The B-mode polarizations, foregrounds and noise sources are described in section \ref{signalnoise}. In section \ref{method}, the method used in this paper is revealed. In section \ref{results}, we show our forecasts for the future experiments. Conclusions and discussion are given in section \ref{discussion}.


\section{Signal, Foregrounds and Noise}\label{signalnoise}

In general, the CMB linear polarizations can be expressed in terms of the spin-weighted spherical harmonics $_{\pm 2}Y_{\ell m}$, namely
\begin{equation}
Q\pm iU=\sum_{\ell m} a^{\pm 2}_{\ell m} ~_{\pm 2}Y_{\ell m}\ ,
\end{equation}
or equivalently, defined by the E- and B-modes as
\begin{eqnarray}
E&=&\sum_{\ell m} a^{E}_{\ell m} Y_{\ell m}\ ,\\
B&=&\sum_{\ell m} a^{B}_{\ell m} Y_{\ell m}\ ,
\end{eqnarray}
where the coefficients are given by
\begin{eqnarray}
a^{E}_{\ell m}&=&-\frac{1}{2}\left(a^{+2}_{\ell m}+a^{-2}_{\ell m}\right)\ ,\\
a^{B}_{\ell m}&=&-\frac{1}{2i}\left(a^{+2}_{\ell m}-a^{-2}_{\ell m}\right)\ .
\end{eqnarray}
In this paper, we are just focused on studying the B-modes which may include the signal of primordial gravitational waves.

In the linear perturbation theory, the primordial B-modes are Gaussian with zero mean, and their variance is given by
\begin{equation}
\langle a^{B}_{\ell m}a^{B\ast}_{\ell^\prime m^{\prime}} \rangle=C^{BB}_{\ell}\delta_{\ell \ell^{\prime}}\delta_{m m^{\prime}}\ ,
\end{equation}
where $\delta$ comes from statistical isotropy.
As conventions, the angular correlation coefficients between B-modes are defined as
\begin{equation}
\tilde{C}_{\ell}=\frac{\ell(\ell+1)}{2\pi}C^{BB}_{\ell}\ ,
\end{equation}
where we have dropped the superscript BB for simplicity. In our study, we generate the power spectrum of primordial B-modes by running the CAMB package \cite{Lewis:1999bs,camb2012}, and we set all the cosmological parameters except $r$ and $n_t$ to the best-fit values of Planck 2015 results \cite{Ade:2015xua}. However, we should find out a reasonable pivot scale $k_p$ such that there is least degeneracy between $r$ and $n_t$ in Eq.~(\ref{tensor power spectrum}).

The foregrounds contaminate the CMB B-mode signal and should be taken into account in the forecast. Actually, we can separate each component of foregrounds by noting that they have very different frequency--dependence. The Galactic synchrotron emission (S) and thermal dust emission (D) are considered in this paper. Their power spectra are given by
\begin{eqnarray}
S_{\ell\nu}&=&\left(W^{S}_{\nu}\right)^{2}C^{S}_{\ell}=\left(W^{S}_{\nu}\right)^{2}A_{S}\left(\frac{\ell}{\ell_{S}}\right)^{\alpha_{S}}\ ,\\
D_{\ell\nu}&=&\left(W^{D}_{\nu}\right)^{2}C^{D}_{\ell}=\left(W^{D}_{\nu}\right)^{2}A_{D}\left(\frac{\ell}{\ell_{D}}\right)^{\alpha_{D}}\ \ ,
\end{eqnarray}
where various parameters can be found in Tab.~\ref{tab:foregroundparameters},
\begin{table}[!hts]
\scriptsize
\centering
\renewcommand{\arraystretch}{1.5}
\begin{tabular}{c|c|c}
 \hline\hline
  Parameters & Synchrotron & Thermal Dust \\
\hline
$A_{72\%}$& $2.1\times10^{-5}$ &  $0.169$ \\
$A_{53\%}$& $2.1\times10^{-5}$ &  $0.065$ \\
$A_{24\%}$& $2.1\times10^{-5}$ &  $0.019$ \\
$A_{11\%}$& $4.2\times10^{-6}$ &  $0.013$ \\
$A_{1\%}$& $4.2\times10^{-6}$ &  $0.006$ \\
$\nu [GHz]$& $65$ &  $353$ \\
$\ell$& $80$ &  $80$ \\
$\alpha$& $-2.6$ &  $-2.42$ \\
$\beta$& $-2.9$ &  $1.59$ \\
$T[K]$& $--$ &  $19.6$ \\
\hline
\end{tabular}
\caption{A list of foreground parameters \cite{Flauger:2014qra,Adam:2014bub,Creminelli:2015oda,Page:2006hz}. Here $A_{f_{sky}}$ denotes the cleanest effective area $f_{sky}$ in the sky and its unit is $\mu K^2$. }
\label{tab:foregroundparameters}
\end{table}
and $W^{S}_{\nu}$ and $W^{D}_{\nu}$ are defined by
\begin{eqnarray}
W^{S}_{\nu}&=&\frac{W^{CMB}_{\nu_{S}}}{W^{CMB}_{\nu}}\left(\frac{\nu}{\nu_{S}}\right)^{\beta_{S}}\ ,\\
W^{D}_{\nu}&=&\frac{W^{CMB}_{\nu_{D}}}{W^{CMB}_{\nu}}\left(\frac{\nu}{\nu_{D}}\right)^{1+\beta_{D}}\frac{e^{h\nu_{D}/k_{B}T}-1}{e^{h\nu/k_{B}T}-1}\ , \\
W^{CMB}_{\nu}&=&\frac{x^2 e^x}{\left(e^x - 1\right)^2}, ~~~~x=\frac{h\nu}{k_B T_{CMB}}\ .
\end{eqnarray}
Here we have rescaled the temperature of Galactic synchrotron emission and thermal dust emission with respect to the CMB temperature. There might be certain correlation between synchrotron emission and dust emission. To account for this, we assume their correlation taking the form $g\sqrt{S_{\ell \nu_i}D_{\ell \nu_j}}$ in the power spectra.

In this paper, we do not consider the systematics which strongly depend on the experimental setups. However, we consider the instrumental white noise which is Gaussian. The power spectrum of the white noise can be expressed as \cite{Knox:1995dq}
\begin{equation}
\mathcal{N}_{\ell}=\frac{\ell(\ell+1)}{2\pi}\delta P^{2}e^{\ell^2 \sigma_b^2}\ ,
\end{equation}
where $\delta P$ denotes the sensitivity for the Stokes parameters Q and U, and $\sigma_b=0.425\theta_{FWHM}$ denotes the beam-size variance. Various instrumental parameters here can be found in Tab.~\ref{tab:instrumentalspecifications} in Appendix.

The gravitational lensing can also limit our ability to detect primordial B-mode polarizations. However, the lensing B-modes and primordial B-modes have the same frequency--dependence, and we cannot separate them as mentioned above. Fortunately, one can reconstruct the lensing potential by considering the CMB data of temperature and E-mode polarizations at small angular scales, and then remove the lensing B-modes away at large angular scales \cite{Knox:2002pe,Kesden:2002ku,Seljak:2003pn,Smith2012}. In this paper, we assume the power of lensing B-modes reduced to $10\%$ of its original value for the CMBPol and COrE experiments. For others, it is marginal to employ delensing \cite{Creminelli:2015oda}. The residual power $\delta C_{\ell}^{lensing}$ of lensing B-modes can be incorporated into the power spectrum of an effective noise, i.e. $\mathcal{N}_\ell\longrightarrow \mathcal{N}_\ell+\delta C_{\ell}^{lensing}$ \cite{Lee:2014cya}.

\section{Likelihood and Fisher matrix}\label{method}

As mentioned above, we will deal with three components of CMB B-modes which originate from primordial gravitational waves, Galactic dust and synchrotron emissions, respectively.
In the ``Component Separation'' (CS) method, the average of log-likelihood can be given by \cite{Creminelli:2015oda}
\begin{equation}\label{likelihood}
\langle\log \mathcal{L}_{BB}\rangle=-\frac{1}{2}\sum_{\ell}(2\ell+1)f_{sky}\left(\log \det \left(\frac{W\mathcal{C}_{\ell}W^{T}+\mathcal{N}_{\ell}}{\bar{W}\mathcal{\bar{C}}_{\ell}\bar{W}^{T}+\mathcal{N}_{\ell}}\right)+\textrm{tr}\left(\frac{\bar{W}\mathcal{\bar{C}}_{\ell}\bar{W}^{T}+\mathcal{N}_{\ell}}{W\mathcal{C}_{\ell}W^{T}+\mathcal{N}_{\ell}}-1\right)\right)\ ,
\end{equation}
where the bar denotes all the parameters fixed to their ``true'' values, and we have used the normalization such that $\langle\log\mathcal{L}_{BB}\rangle=0$ for $\mathcal{\bar{C}}_\ell=\mathcal{C}_\ell$ and $\bar{W}=W$. Here $W$ denotes the frequency--dependence of each component of CMB B-modes, which is a $3\times N$ matrix with a row $(1,W_{\nu_i}^D,W_{\nu_i}^{S})$. $N$ denotes the number of frequency channels for each experiment. $\mathcal{C}_\ell$ is the covariance matrix of the amplitudes of three components. $f_{sky}$ stands for the effective area of the sky in Tab.~\ref{tab:instrumentalspecifications}, where each experiment observes.

In our consideration, the average of log-likelihood is a function of several parameters $\mathbf{p}$ which are $(r,n_t,A_D,A_S,\beta_D,\beta_S,g)$. We assume Gaussian priors for $A_D$, $A_S$, $\beta_D$ and $\beta_S$ with the variance of $50\%$, $50\%$, $15\%$ and $10\%$. Based on Eq.~(\ref{likelihood}), the Fisher matrix is defined by
\begin{equation}
F_{ij}=-\frac{\partial^2\langle\log\mathcal{L}_{BB}\rangle}{\partial p_i \partial p_j}\mid_{\mathbf{p}=\mathbf{\bar{p}}}\ ,
\end{equation}
where $\bar{\mathbf{p}}$ denote ``true'' values of the parameters $\mathbf{p}$. In our fiducial model, we set $\bar{g}=0.5$ and $\bar{n}_t=0$.
The minimum error on the parameter $p_i$ is given by the Cramer-Rao bound, i.e.
\begin{equation}
\sigma^2_{p_i}\geqslant\left(F^{-1}\right)_{ii}\ .
\end{equation}
In this paper, we will forecast the future sensitivity on both $r$ and $n_t$ simultaneously, since primordial gravitational waves are not detected until now. There could be certain correlation between $r$ and $n_t$, or equivalently, the $(r,n_t)$ confidence ellipse has a tilt. Thus we should find a pivot scale $k_p$ to make their correlation to be least. In other words, we should find the pivot scale such that $\left(F^{-1}\right)_{rn_t}=0$, which minimizes the constraint on $r$.

\section{Analysis and Results}\label{results}

For the future ground-based experiments, we consider the CMB multipoles of the range $[30,150]$ for  Keck/BICEP3, Simons Array and SPT-3G, while $[2,150]$ for AdvACT and CLASS. The reason is that AdvACT and CLASS cover a larger fraction of the sky. We forecast on the future sensitivity on $r$ and $n_t$ by using the Fisher matrix. The $1\sigma$ errors on $r$ and $n_t$ can be found in Tab.~\ref{tab:ground01}. We also list the pivot scale $k_p[\rm{Mpc}^{-1}]$ at which there is no tilt for the $(r,n_t)$ confidence ellipse.
\begin{table}[htbp]
\begin{center}
\begin{tabular}{ c| c| c| c| c}
\hline\hline
& {\bf r} & {$\boldsymbol{ \sigma_r}$} & $\boldsymbol{ \sigma_{n_{t}}}$ & $\boldsymbol{ k_{p}[\rm{Mpc^{-1}}]}$ \\
\hline
\multicolumn{1}{ c| }{} 	                                              	 & 0.1 	& $5.3\times 10^{-3}$	& $9.7\times 10^{-2}$	& $8.4\times 10^{-3}$	\\
\multicolumn{1}{ c| }{} 							& 0.05 	& $4.4\times 10^{-3}$	& $1.2\times 10^{-1}$	& $7.5\times 10^{-3}$ 				 \\
\multicolumn{1}{ c| }{\multirow{2}{*}{{\bf AdvACT}}} 				 & 0.02 	& $3.6\times 10^{-3}$	& $1.6\times 10^{-1}$	& $5.2\times 10^{-3}$ 				\\
\multicolumn{1}{ c| }{} 							& 0.01 	& $3.1\times 10^{-3}$	& $2.3\times 10^{-1}$	 & $3.4\times 10^{-3}$ 				\\
\multicolumn{1}{ c| }{} 							& 0.005 	& $2.6\times 10^{-3}$	& $3.6\times 10^{-1}$	& $2.1\times 10^{-3}$ 				 \\
\multicolumn{1}{ c| }{} 							& 0.002 	& $2.1\times 10^{-3}$	& $7.5\times 10^{-1}$	& $1.3\times 10^{-3}$ 				 \\
\hline
\multicolumn{1}{ c| }{} 		                                        & 0.1 	& $6.7\times 10^{-3}$	& $9.0\times 10^{-2}$	& $6.3\times 10^{-3}$ 		\\
\multicolumn{1}{ c| }{} 							& 0.05 	& $5.5\times 10^{-3}$	& $1.1\times 10^{-1}$	& $4.7\times 10^{-3}$ 				 \\
\multicolumn{1}{ c| }{\multirow{2}{*}{{\bf CLASS}}} 				 & 0.02 	& $4.1\times 10^{-3}$	& $1.6\times 10^{-1}$	& $2.4\times 10^{-3}$ 				\\
\multicolumn{1}{ c| }{} 							& 0.01 	& $3.1\times 10^{-3}$	& $2.5\times 10^{-1}$	 & $1.3\times 10^{-3}$ 				\\
\multicolumn{1}{ c| }{} 							& 0.005 	& $2.2\times 10^{-3}$	& $4.2\times 10^{-1}$	& $8.6\times 10^{-4}$ 				 \\
\multicolumn{1}{ c| }{} 							& 0.002 	& $1.5\times 10^{-3}$	& $9.3\times 10^{-1}$	& $5.7\times 10^{-4}$ 				 \\
\hline
\multicolumn{1}{ c| }{} 	                                                & 0.1 	& $2.3\times 10^{-2}$ 	& $1.2$	& $9.2\times 10^{-3}$\\
\multicolumn{1}{ c| }{} 							& 0.05 	& $1.7\times 10^{-2}$	& $1.8$	& $8.9\times 10^{-3}$ 				 \\
\multicolumn{1}{ c| }{\multirow{2}{*}{{\bf Keck/BICEP3}}} 			 & 0.02 	& $1.4\times 10^{-2}$	& $3.6$	& $8.7\times 10^{-3}$ 				 \\
\multicolumn{1}{ c| }{} 							& 0.01 	& $1.3\times 10^{-2}$	& $6.6$	& $8.7\times 10^{-3}$ 				 \\
\multicolumn{1}{ c| }{} 							& 0.005 	& $1.2\times 10^{-2}$	& $12.5$	& $8.6\times 10^{-3}$ 				 \\
\multicolumn{1}{ c| }{} 							& 0.002 	& $1.2\times 10^{-2}$	& $30.4$	& $8.6\times 10^{-3}$ 				 \\
\hline
\multicolumn{1}{ c| }{} 	                                                & 0.1 	& $1.7\times 10^{-2}$	& $7.0\times 10^{-1}$	& $9.3\times 10^{-3}$\\
\multicolumn{1}{ c| }{} 							& 0.05 	& $1.5\times 10^{-2}$	& $1.3$	& $9.3\times 10^{-3}$ 				 \\
\multicolumn{1}{ c| }{\multirow{2}{*}{{\bf Simons Array}}} 							 & 0.02 	 & $1.4\times 10^{-2}$	& $3.0$	& $9.2\times 10^{-3}$ 				 \\
\multicolumn{1}{ c| }{} 			& 0.01 	& $1.4\times 10^{-2}$	 & $5.9$	& $9.2\times 10^{-3}$ 				\\
\multicolumn{1}{ c| }{} 							& 0.005 	& $1.4\times 10^{-2}$	& $11.7$	& $9.2\times 10^{-3}$ 				 \\
\multicolumn{1}{ c| }{} 							& 0.002 	& $1.4\times 10^{-2}$	& $29.1$	& $9.2\times 10^{-3}$ 				 \\
\hline
\multicolumn{1}{ c| }{} 		                                        & 0.1 	& $8.8\times 10^{-3}$	& $4.3\times 10^{-1}$	& $9.4\times 10^{-3}$\\
\multicolumn{1}{ c| }{} 							& 0.05 	& $6.6\times 10^{-3}$	& $6.4\times 10^{-1}$	& $9.2\times 10^{-3}$ 				 \\
\multicolumn{1}{ c| }{\multirow{2}{*}{{\bf SPT-3G}}} 							 & 0.02 	 & $5.3\times 10^{-3}$	& $1.3$	& $9.1\times 10^{-3}$ 				 \\
\multicolumn{1}{ c| }{} 				& 0.01 	& $4.8\times 10^{-3}$	& $2.3$	& $9.1\times 10^{-3}$ 				\\
\multicolumn{1}{ c| }{} 							& 0.005 	& $4.5\times 10^{-3}$	& $4.4$	 & $9.0\times 10^{-3}$ 				 \\
\multicolumn{1}{ c| }{} 							& 0.002 	& $4.4\times 10^{-3}$	& $10.6$	& $9.0\times 10^{-3}$ 				 \\
\hline
\end{tabular}
\caption{$1\sigma$ errors on $r$ and $n_t$ for future ground-based experiments. The pivot scale $k_p[\rm{Mpc^{-1}}]$ is also listed, at which there is no correlation between $r$ and $n_t$ and then the ($r$,$n_t$) confidence ellipse has no tilt. }
\label{tab:ground01}
\end{center}
\end{table}
We found that the ground-based experiments can probe the tensor-to-scalar ratio to the level $r\simeq0.01$, which is consistent with previous estimates. The $1\sigma$ error on the tensor tilt is to the level $\sigma_{n_t}\sim0.1$ when $2\times10^{-3}<r<0.1$ for AdvACT and CLASS, while $\sigma_{n_t}\gtrsim 1$ for other three experiments.

For the future balloon-borne experiments, we consider the CMB multipoles of the range $[30,150]$. The $1\sigma$ errors on $r$ and $n_t$ and the pivot scale $k_p$ can be found in Tab.~\ref{tab:balloon01}.
\begin{table}[htbp]
\begin{center}
\begin{tabular}{ c| c| c| c| c}
\hline\hline
& {\bf r} & {$\boldsymbol{ \sigma_r}$} & $\boldsymbol{ \sigma_{n_{t}}}$ & $\boldsymbol{ k_{p}[\rm{Mpc^{-1}}]}$ \\
\hline
\multicolumn{1}{ c| }{} 							& 0.1 	& $2.3\times 10^{-2}$	& $1.2$	& $9.1\times 10^{-3}$	\\
\multicolumn{1}{ c| }{} 							& 0.05 	& $1.7\times 10^{-2}$	& $1.9$	& $8.8\times 10^{-3}$ 				 \\
\multicolumn{1}{ c| }{\multirow{2}{*}{{\bf EBEX}}} 							 & 0.02 	& $1.3\times 10^{-2}$	& $3.7$	& $8.6\times 10^{-3}$ 				 \\
\multicolumn{1}{ c| }{} 				& 0.01 	& $1.2\times 10^{-2}$	& $6.8$	& $8.4\times 10^{-3}$ 				\\
\multicolumn{1}{ c| }{} 							& 0.005 	& $1.1\times 10^{-2}$	& $12.9$	& $8.4\times 10^{-3}$ 				 \\
\multicolumn{1}{ c| }{} 							& 0.002 	& $1.0\times 10^{-2}$	& $31.4$	& $8.4\times 10^{-3}$ 				 \\
\hline
\multicolumn{1}{ c| }{} 							& 0.1 	& $1.8\times 10^{-2}$	& $1.0$	& $8.6\times 10^{-3}$ 		\\
\multicolumn{1}{ c| }{} 							& 0.05 	& $1.5\times 10^{-2}$	& $1.8$	& $8.4\times 10^{-3}$ 				 \\
\multicolumn{1}{ c| }{\multirow{2}{*}{{\bf Spider}}} 							 & 0.02 	 & $1.4\times 10^{-2}$	& $4.1$	& $8.3\times 10^{-3}$ 				 \\
\multicolumn{1}{ c| }{} 				& 0.01 	& $1.3\times 10^{-2}$	& $7.9$	& $8.3\times 10^{-3}$ 				\\
\multicolumn{1}{ c| }{} 							& 0.005 	& $1.3\times 10^{-2}$	& $15.4$	& $8.3\times 10^{-3}$ 				 \\
\multicolumn{1}{ c| }{} 							& 0.002 	& $1.3\times 10^{-2}$	& $38.2$	& $8.3\times 10^{-3}$ 				 \\
\hline
\end{tabular}
\caption{$1\sigma$ errors on $r$ and $n_t$ for future balloon-borne experiments. The pivot scale $k_p[\rm{Mpc^{-1}}]$ is also listed correspondingly. }
\label{tab:balloon01}
\end{center}
\end{table}
We find that EBEX and Spider can marginally probe the tensor-to-scalar ratio to the level $r\simeq 0.02$, which is consistent with previous estimates. The $1\sigma$ error on the tensor tilt is $\sigma_{n_t}>1$ when $2\times10^{-3}<r<0.1$.

For the future satellite experiments, we consider the CMB multipoles of the range $[2,300]$ for CMBPol and COrE while $[2,150]$ for LiteBIRD. The $1\sigma$ errors on $r$ and $n_t$ and the pivot scale $k_p$ can be found in Tab.~\ref{tab:satellite01}.
\begin{table}[htbp]
\begin{center}
\begin{tabular}{ c| c| c| c| c}
\hline\hline
& {\bf r} & {$\boldsymbol{ \sigma_r}$} & $\boldsymbol{ \sigma_{n_{t}}}$ & $\boldsymbol{ k_{p}[\rm{Mpc^{-1}}]}$ \\
\hline
\multicolumn{1}{ c| }{} 							& 0.1 	& $1.2\times 10^{-3}$	& $3.0\times 10^{-2}$	& $1.1\times 10^{-2}$	 \\
\multicolumn{1}{ c| }{} 							& 0.05 	& $7.1\times 10^{-4}$	& $3.8\times 10^{-2}$	& $1.0\times 10^{-2}$ 				 \\
\multicolumn{1}{ c| }{\multirow{2}{*}{{\bf CMBPol}}} 							 & 0.02 	 & $3.8\times 10^{-4}$	& $5.1\times 10^{-2}$	& $9.0\times 10^{-3}$ 				\\
\multicolumn{1}{ c| }{} 				& 0.01 	& $2.5\times 10^{-4}$	& $6.2\times 10^{-2}$	 & $8.4\times 10^{-3}$ 				 \\
\multicolumn{1}{ c| }{} 							& 0.005 	& $1.8\times 10^{-4}$	& $7.5\times 10^{-2}$	& $7.9\times 10^{-3}$ 				 \\
\multicolumn{1}{ c| }{} 							& 0.002 	& $1.4\times 10^{-4}$	& $9.7\times 10^{-2}$	& $7.2\times 10^{-3}$ 				 \\
\hline
\multicolumn{1}{ c| }{} 							& 0.1 	& $1.3\times 10^{-3}$	& $3.4\times 10^{-2}$	& $1.1\times 10^{-2}$ 		 \\
\multicolumn{1}{ c| }{} 							& 0.05 	& $8.2\times 10^{-4}$	& $4.4\times 10^{-2}$	& $9.7\times 10^{-3}$ 				 \\
\multicolumn{1}{ c| }{\multirow{2}{*}{{\bf COrE}}} 							 & 0.02 	& $4.6\times 10^{-4}$	& $5.8\times 10^{-2}$	& $8.6\times 10^{-3}$ 				\\
\multicolumn{1}{ c| }{} 				& 0.01 	& $3.3\times 10^{-4}$	& $7.0\times 10^{-2}$	 & $8.1\times 10^{-3}$ 				 \\
\multicolumn{1}{ c| }{} 							& 0.005 	& $2.5\times 10^{-4}$	& $8.3\times 10^{-2}$	& $7.5\times 10^{-3}$ 				 \\
\multicolumn{1}{ c| }{} 							& 0.002 	& $2.1\times 10^{-4}$	& $1.1\times 10^{-1}$	& $6.3\times 10^{-3}$ 				 \\
\hline
\multicolumn{1}{ c| }{} 							& 0.1 	& $1.8\times 10^{-3}$	& $5.3\times 10^{-2}$	& $8.9\times 10^{-3}$ 		 \\
\multicolumn{1}{ c| }{} 							& 0.05 	& $1.1\times 10^{-3}$	& $6.1\times 10^{-2}$	& $8.5\times 10^{-3}$ 				 \\
\multicolumn{1}{ c| }{\multirow{2}{*}{{\bf LiteBIRD}}} 							 & 0.02 	 & $7.3\times 10^{-4}$	& $7.5\times 10^{-2}$	& $8.0\times 10^{-3}$ 				\\
\multicolumn{1}{ c| }{} 				& 0.01 	& $5.8\times 10^{-4}$	& $8.8\times 10^{-2}$	 & $7.1\times 10^{-3}$ 				 \\
\multicolumn{1}{ c| }{} 							& 0.005 	& $4.9\times 10^{-4}$	& $1.1\times 10^{-1}$	& $6.2\times 10^{-3}$ 				 \\
\multicolumn{1}{ c| }{} 							& 0.002 	& $4.4\times 10^{-4}$	& $1.7\times 10^{-1}$	& $5.3\times 10^{-3}$ 				 \\
\hline
\end{tabular}
\caption{$1\sigma$ errors on $r$ and $n_t$ for future satellite experiments. The pivot scale $k_p[\rm{Mpc^{-1}}]$ is also listed correspondingly. }
\label{tab:satellite01}
\end{center}
\end{table}
The delensing has been taken into account for CMBPol and COrE. We find that the satellite experiments can well probe the tensor-to-scalar ratio even to the level $r\simeq2\times10^{-3}$, which is also consistent with previous estimates. The $1\sigma$ error on the tensor tilt $n_t$ runs from $\sim0.1$ to $\sim0.01$ when $r$ runs from $2\times10^{-3}$ to $0.1$. Thus the future satellite experiments provide the highest sensitivity on the tensor tilt $n_t$ by contrast to the future ground-based and balloon-borne experiments.

\begin{figure}[htbp]
\centering
\includegraphics[scale=1]{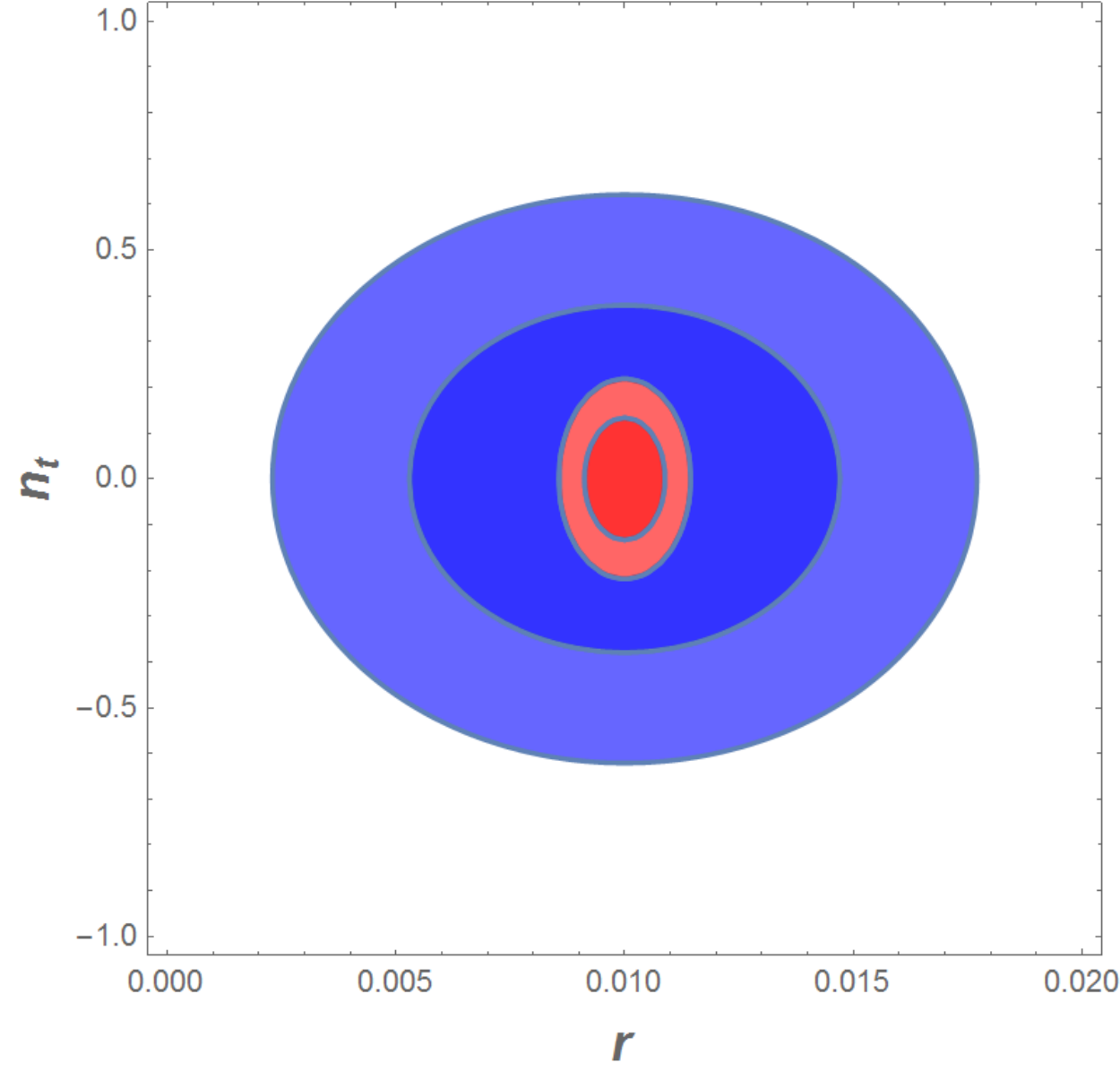}
\caption{The ($r$, $n_t$) confidence ellipses of $1\sigma$ and $2\sigma$ C.L. for CLASS (blue) and LiteBIRD (red). The fiducial parameters are given by $r=0.01$ and $n_t=0$.}
\label{fig:rntplot}
\end{figure}

We plot the ($r$, $n_t$) confidence ellipses of $1\sigma$ and $2\sigma$ C.L. for the CLASS and LiteBIRD experiments in Fig.~\ref{fig:rntplot}.
Here the fiducial parameters are chosen as $r=0.01$ and $n_t=0$. AdvACT and CLASS denote the best sensitivity on both $r$ and $n_t$ in the five future ground-based experiments. Unfortunately, they can not still compare with the future satellite experiments such as LiteBIRD. Even though the satellite experiments give the best sensitivity on $n_t$, it is impossible for them to test the consistent relation, i.e. $n_t=-r/8$. By considering the BKP constraint $r_{0.05}<0.12$ at $95\%$ C.L., we deduce $n_t<0.015$ which is much smaller than the $1\sigma$ errors on $n_t$ in Tab.~\ref{tab:satellite01}.


\section{Discussion}\label{discussion}

In this paper, we forecasted the sensitivity on the tensor tilt $n_t$ for future ground-based, balloon-borne and satellite experiments of B-mode polarizations. We focused on the scale-invariant power spectrum $n_t=0$ which approximately corresponds to the canonical single-field slow-roll inflation models. For the tensor tilt, the satellite experiments can give a good sensitivity on $n_t$ even for $r\simeq2\times10^{-3}$. Specifically, the $1\sigma$ error on $n_t$ varies from $\sim0.1$ to $\sim0.01$ when $r$ varies from $\sim0.001$ to $\sim0.1$. By contrast, the ground-based and balloon-borne experiments give much worse sensitivity on both $r$ and $n_t$. Furthermore, our results did not change significantly with respect to previous forecasts on the future sensitivity of tensor-to-scalar ratio $r$ in Ref.~\cite{Creminelli:2015oda}. The reason is that we just considered the scale-invariant tensor tilt $n_t=0$ in our fiducial model. By considering the BKP constraint on $r$, we see that it is impossible for these future experiments to test the consistent relation $n_t=-r/8$ in the canonical single-field slow-roll inflation models.

\vspace{0.5 cm}
\noindent{\large \bf Acknowledgments}

Q.G.H. and S.W. are supported by grants from NSFC (grant NO. 11322545, 11335012 and 11575271). W.Z. is supported by Project 973 under Grant No. 2012CB821804, by NSFC No. 11173021, 11322324, 11421303 and project of KIP and CAS.

\newpage

\begin{appendix}

\section{Instrumental specifications}
\begin{table}[htbp]
\tiny
\begin{center}
\begin{tabular}{ c| c| c| c| c}
\hline\hline
Experiments & {$f_{sky}[\%]$} & $\nu[GHz]$ & $\theta_{FWHW}[^\prime]$ & $\delta P [\mu K^\prime]$ \\
\hline
\multicolumn{1}{ c| }{\multirow{2}{*}{{AdvACT}}} 							 & $50$ 	& $90$	 & $2.2$	& $7.8$ 				\\
\multicolumn{1}{ c| }{} 							& $50$ 	& $150$	& $1.3$	& $6.9$ 				 \\
\multicolumn{1}{ c| }{} 				& $50$ 	& $230$	& $0.9$	& $25$ 				\\
\hline
\multicolumn{1}{ c| }{\multirow{2}{*}{{CLASS}}} 							 & $70$ 	& $38$	 & $90$	& $39$ 		\\
\multicolumn{1}{ c| }{} 							& $70$ 	& $93$	& $40$	& $13$ 				 \\
\multicolumn{1}{ c| }{} 							& $70$ 	& $148$	& $24$	& $15$ 				 \\
\multicolumn{1}{ c| }{} 							& $70$ 	& $217$	& $18$	& $43$ 				 \\
\hline
\multicolumn{1}{ c| }{\multirow{2}{*}{{Keck/BICEP3}}} 							 & $1$ 	& $95$	 & $30$	& $9.0$ 				\\
\multicolumn{1}{ c| }{} 							& $1$ 	& $150$	& $30$	& $2.3$ 				 \\
\multicolumn{1}{ c| }{} 				& $1$ 	& $220$	& $30$	& $10$ 				\\
\hline
\multicolumn{1}{ c| }{\multirow{2}{*}{{Simons Array}}} 							 & $20$ 	 & $90$	 & $5.2$	& $15.2$ 				\\
\multicolumn{1}{ c| }{} 							& $20$ 	& $150$	& $3.5$	& $12.3$ 				 \\
\multicolumn{1}{ c| }{} 				& $20$ 	& $220$	& $2.7$	& $23.6$ 				 \\
\hline
\multicolumn{1}{ c| }{\multirow{2}{*}{{SPT-3G}}} 							 & $6$ 	& $95$	 & $1$	& $6.0$ 				\\
\multicolumn{1}{ c| }{} 							& $6$ 	& $150$	& $1$	& $3.5$ 				 \\
\multicolumn{1}{ c| }{} 				& $6$ 	& $220$	& $1$	& $6.0$ 				 \\
\hline\hline
\multicolumn{1}{ c| }{\multirow{2}{*}{{EBEX}}} 							 & $1$ 	& $150$	& $8$	 & $5.8$ 				\\
\multicolumn{1}{ c| }{} 							& $1$ 	& $250$	& $8$	& $17$ 				 \\
\multicolumn{1}{ c| }{} 				& $1$ 	& $410$	& $8$	& $150$ 				 \\
\hline
\multicolumn{1}{ c| }{\multirow{2}{*}{{Spider}}} 							 & $7.5$ 	& $94$	 & $49$	& $17.8$ 				\\
\multicolumn{1}{ c| }{} 							& $7.5$ 	& $150$	& $30$	& $13.6$ 				 \\
\multicolumn{1}{ c| }{} 				& $7.5$ 	& $280$	& $17$	& $52.6$ 				 \\
\hline\hline
\multicolumn{1}{ c| }{} 							& $70$ 	& $30$	 & $26$	& $19.2$ 				 \\
\multicolumn{1}{ c| }{\multirow{2}{*}{{CMBPol}}} 							 & $70$ 	& $45$	 & $17$	& $8.3$ 				 \\
\multicolumn{1}{ c| }{} 							& $70$ 	& $70$	& $11$	& $4.2$ 				 \\
\multicolumn{1}{ c| }{} 							& $70$ 	& $100$	& $8$	& $3.2$ 				 \\
\multicolumn{1}{ c| }{} 							& $70$ 	& $150$	& $5$	& $3.1$ 				 \\
\multicolumn{1}{ c| }{} 							& $70$ 	& $220$	& $3.5$	& $4.8$ 				 \\
\multicolumn{1}{ c| }{} 							& $70$ 	& $340$	& $2.3$	& $21.6$ 		 \\
\hline
\multicolumn{1}{ c| }{} 							& $70$ 	& $45$	& $23$	& $9.1$ 		 \\
\multicolumn{1}{ c| }{\multirow{2}{*}{{COrE}}} 							 & $70$ 	& $75$	 & $14$	 & $4.7$ 				\\
\multicolumn{1}{ c| }{} 							& $70$ 	& $105$	& $10$	& $4.6$ 				 \\
\multicolumn{1}{ c| }{} 							& $70$ 	& $135$	& $7.8$	& $4.6$ 				 \\
\multicolumn{1}{ c| }{} 							& $70$ 	& $165$	& $6.4$	& $4.6$ 				 \\
\multicolumn{1}{ c| }{} 							& $70$ 	& $195$	& $5.4$	& $4.5$ 				 \\
\multicolumn{1}{ c| }{} 							& $70$ 	& $225$	& $4.7$	& $4.6$ 				 \\
\multicolumn{1}{ c| }{} 							& $70$ 	& $255$	& $4.1$	& $10.5$ 				 \\
\multicolumn{1}{ c| }{} 							& $70$ 	& $285$	& $3.7$	& $17.4$ 				 \\
\multicolumn{1}{ c| }{} 							& $70$ 	& $315$	& $3.3$	& $46.6$ 				 \\
\multicolumn{1}{ c| }{} 							& $70$ 	& $375$	& $2.8$	& $119$ 				 \\
\hline
\multicolumn{1}{ c| }{} 							& $70$ 	& $60$	& $32$	& $10.3$ 		 \\
\multicolumn{1}{ c| }{\multirow{2}{*}{{LiteBIRD}}} 							 & $70$ 	& $78$	 & $58$	& $6.5$ 				\\
\multicolumn{1}{ c| }{} 							& $70$ 	& $100$	& $45$	& $4.7$ 				 \\
\multicolumn{1}{ c| }{} 							& $70$ 	& $140$	& $32$	& $3.7$ 				 \\
\multicolumn{1}{ c| }{} 							& $70$ 	& $195$	& $24$	& $3.1$ 				 \\
\multicolumn{1}{ c| }{} 							& $70$ 	& $280$	& $16$	& $3.8$ 				 \\
\hline
\end{tabular}
\caption{Instrumental specifications of future ground-based, balloon-borne and satellite experiments \cite{Calabrese:2014gwa,Reichborn-Kjennerud001,Ogburn:2012ma,Lee001,Benson:2014qhw,Fraisse:2011xz,Rahlin:2014rja,Baumann:2008aq,Bouchet:2011ck,Matsumura:2013aja}. Here $\delta P=\sigma_{pix}\theta_{FWHM}$. }
\label{tab:instrumentalspecifications}
\end{center}
\end{table}

\end{appendix}

\newpage

\end{document}